\documentclass[11pt]{revtex4}
\usepackage{epsfig}
\usepackage{amsfonts}
\usepackage{amsmath}
\usepackage{amssymb}
\usepackage{graphicx}
\usepackage{mathptmx}

\def\be{\begin{eqnarray}}
\def\en{\end{eqnarray}}
\def\non{\nonumber}
\def\pr{{Phys. Rev.}~}

\def\B{{BR}}
\def\ra{\rangle}

\def\sl{\!\!\!\!\slash}
\def\ov{\overline}
\begin{document}
\title{Implications of Family Nonuniversal $Z^\prime$ Model on  $B \to K_0^*\pi$ Decays}
\author{Ying Li\footnote{Email:liying@ytu.edu.cn},
\,\,\,\,\,\,\,Xiao-Jiao Fan,
\,\,\,\,\,\,\,Juan Hua,
\,\,\,\,\,\,\,En-Lei Wang
}
\date{\today}
\affiliation{Department of Physics, Yantai University, Yantai 264-005, China}
\begin{abstract}
Within the QCD factorization formalism, we study the possible impacts of
the nonuniversal $Z^\prime$ model, which provides a flavor-changing
neutral current at the tree level, on rare decays $B \to K_0^*\pi$.
Under two different scenarios (S1 and S2) for identifying the scalar
meson $K_0^*(1430)$, the branching ratios, $CP$ asymmetries, and
isospin asymmetries are calculated in both the standard model (SM)
and the family nonuniversal $Z^\prime$ model. We find that the
branching ratios and $CP$ asymmetries are sensitive to weak
annihilation. In the SM, with $\rho_A=1$ and $\phi_A\in[-30^\circ,
30^\circ]$, the branching ratios of S1 (S2) are  smaller (larger)
than the experimental data. Adding the contribution of the $Z^\prime$
boson in two different cases (Case-I and Case-II), for S1, the
branching ratios are still far away from experiment. For S2, in
Case-II, the branching ratios become smaller and can accommodate the
data; in Case-I, although the center values are enhanced, they can
also explain the data with large uncertainties. Similar
conclusions are also reached for $CP$ asymmetries. Our results
indicate that S2 is more favored  than S1, even after considering new
physics effects. Moreover, if there exists a nonuniversal
$Z^\prime$ boson, Case-II is preferred. All results can be
tested in the LHC-b experiment and forthcoming super-B
factory.
\end{abstract}
 \maketitle
\section{Introduction}\label{sec:1}
Recently, with rich events in two $B$ factories,  measurements of
$B$ meson nonleptonic  charmless decays involving scalar mesons have
become available. Among these decays, the processes $B \to K_0^*\pi$
are attractive since they are dominantly induced by the flavor-changing
neutral current (FCNC) transition $b \to s\bar qq~(q=u,d,s)$. Such a
transition forbidden at the tree level in the Standard Model (SM)
is expected to be an excellent ground for testing SM and searching
for new physics (NP) beyond SM. Therefore, many similar decay
modes induced by FCNC have been explored widely in the literatures,
such as $B\to K\pi, K\eta^{(\prime)}, \phi K^{(*)}$. The recent
reviews can be found, for example, in Ref.~\cite{Cheng:2010yv}. For
the concerned decay modes $B \to K_0^*\pi$, the latest world averaged
branching ratios from Heavy Flavor Average Group~\cite{Asner:2010qj} are listed as:
\begin{eqnarray}\label{data}
\B(B^+\to K^{*0}_0(1430)\pi^+) &=&  (45.1\pm 6.3)\times 10^{-6};\nonumber \\
\B(B^0\to K^{*+}_0(1430)\pi^-) &=&  (33.5^{+3.9}_{-3.8})\times 10^{-6};\nonumber\\
\B(B^0\to K^{*0}_0(1430)\pi^0) &=&  (11.7^{+4.2}_{-3.8})\times 10^{-6}.
\end{eqnarray}
Direct $CP$ asymmetries of above decays have also been measured
recently by BaBar  and Belle experiments, which will be shown in
Sec.~\ref{sec:4}. As direct $CP$ violation is sensitive to the strong
phase involved in the decay process, the comparison between theory
and experiment will offer us information on the strong phases
necessary for producing the measured direct $CP$ asymmetries.
Comparing the predicted results of the SM~\cite{Cheng:2005nb} with experimental
data, ie. Eq.(\ref{data}), we notice that the theoretical results cannot accommodate the data well even with large uncertainties. So, it is worth while to explore whether some new physics models could explain the data.

When discussing the $B$ meson non-leptonic charmless decays, the
hadronic matrix elements  are required. In the past few years,
several novel methods have been proposed to study matrix elements
related to exclusive hadronic $B$ decays, such as naive
factorization (NF) \cite{Wirbel:1985ji}, generalized
factorization \cite{Ali:1998eb}, the perturbative QCD method (pQCD)
\cite{Lu:2000em}, QCD factorization (QCDF) \cite{Beneke:1999br},
the soft collinear effective theory (SCET) \cite{Bauer:2001cu}, and
so on.  Among these approaches, QCDF based on collinear
factorization is a systematic framework to compute these matrix
elements from QCD theory, and it holds in the heavy quark
limit $m_b \to \infty$ and the heavy quark symmetry. Thus, we shall use QCDF approach in the
following calculations.

Although the study of scalar meson spectrum has been an interesting
topic   for a long time, the underlying structure of the light
scalar meson is still controversial until now. In the
literature, there are many schemes for the classification of them.
Here we present two typical scenarios to describe the scalar mesons
\cite{Tornqvist:1982yv}. Scenario-1 (S1) is the naive 2-quark
model: the nonet mesons below 1 GeV are treated as the lowest lying
states, and the ones near 1.5 GeV are the first orbitally excited
states. In scenario-2 (S2), the nonet mesons near 1.5 GeV are
regarded as the lowest lying states, while the mesons below 1 GeV
may be viewed as exotic states beyond the two-quark model. Since the
mass of $K^{*}_0(1430)$ is very near 1.5 GeV, thus it should be
composed by two quarks in both S1 and S2, but the decay constants
and distribution amplitudes are different in the different
scenarios. Under above pictures, the two body nonleptonic $B$ decays
involving scalar mesons have been explored in both  QCDF
\cite{Cheng:2005ye,Cheng:2005nb,Cheng:2007st} and pQCD approaches
\cite{Chen:2002si,Chen:2005cx,Wang:2006ria,shen,Kim:2009dg}.

As stated before, $B \to K_0^*\pi$ decays are dominantly induced by FCNC
$b \to s\bar qq$ transition, hence they are sensitive to new
physics contributions even if they are suppressed by a large mass
parameter which characterizes the new physics scale. To search for
signals of NP, a model independent analysis is not suitable for the
current status. It is the purpose of this work to show that a  new
physics effect of similar size can be obtained from some models with
an extra $Z^\prime$ boson. $Z^\prime$ bosons are known to naturally
exist in certain well-motivated extensions of the SM, such as the
string theory \cite{String}, the grand unified theories \cite{GUTs},
the little Higgs models\cite{LH}, light U-boson model \cite{Uboson},
by adding additional $U(1)^\prime$ gauge symmetry. Among those
models,  a well-motivated $Z^\prime$ model for low energy systems is
the so-called family non-universal $Z^\prime$ model, where the
$Z^\prime$ couplings are affected by fermion mixing and are not
diagonal in the mass basis. Non-trivial flavor-changing neutral
current (FCNC) effects at the tree level mediated by the $Z^\prime$
therefore are induced, which play an important role in explaining
the $CP$ asymmetries in the current high energy experiments by
introducing new weak phases. The effects of $Z^\prime$ boson in $B$
sector have been investigated in a number of papers
\cite{Langacker:2000ju,Barger:2009hn,Cheung:2006tm,Chang:2009wt,Langacker:2008yv}.
In this work, we will show the implications of the family
nonuniversal $Z^\prime$ model on  $B \to K_0^*\pi$ decays.

The layout of this paper is as follows. In Sec.\ref{sec:2}, we
firstly present  the formulaes of $B \to K_0^*\pi$ in the SM within the
QCDF approach, involving the effective Hamiltonian and the
amplitudes. In Sec.\ref{sec:3}, we specify our flavor-changing
$Z^\prime$ model, and how the effective Hamiltonian responsible for
hadronic $B$ decays is modified. The numerical results and
discussions are given in Sect.\ref{sec:4}.The conclusions are
presented in the final section.

\section{Calculation in the Standard Model}\label{sec:2}
In the two-quark picture of S1 and S2, the two kinds of decay constants of scalar meson $S$ are defined by:
\begin{eqnarray}
\langle S(p)|\bar q_2\gamma_\mu q_1|0\ra=f_Sp_\mu, \,\,\,\,\,\,\,\,\,\,\,
\langle S(p)|\bar q_2 q_1|0\ra=m_S\bar {f_S}.
\end{eqnarray}
The vector decay constant $f_S$ and the scale-dependent scalar decay constant $\bar f_S$ are related by equations of motion
\begin{eqnarray} \label{eq:EOM}
 \mu_Sf_S=\bar f_S, \qquad\quad{\rm with}~~\mu_S=\frac{m_S}{m_2(\mu)-m_1(\mu)},
\end{eqnarray}
where $m_{2}$ and $m_{1}$ are the running current quark masses.
Therefore, contrary to  the case  of pseudoscalar one, the vector
decay constant of the scalar meson, namely, $f_S$, will vanish in
the SU(3) limit. In other words, the vector decay constant of $K_0^*(1430)$
is fairly  small.

As for the scalar meson wave function, the twist-2
and twist-3 light-cone distribution amplitudes (LCDAs) for different
components could be combined into a single matrix element:
\begin{eqnarray}
\langle K_0^{*+}(p)|\bar{u}_{\beta}(z)s_{\alpha}(0)|0\rangle
=\frac{1}{\sqrt{6}}\int^1_0dxe^{ixp \cdot z}\bigg\{ p\sl
\phi_{K^{*+}_0}(x) + m_S\phi^S_{K^{*+}_0}(x)+\frac{1}{6}
m_S\sigma_{\mu\nu}p^{\mu}z^{\nu}\phi^{\sigma}_{K^{*+}_0}(x)\bigg\}_{\alpha\beta}.
\end{eqnarray}
The distribution amplitudes $\phi_{K^*_0}(x)$, $\phi^S_{K^*_0}(x)$, and $\phi^{\sigma}_{K^*_0}(x)$ are normalized as:
\begin{eqnarray}
\int^1_0dx\phi_{K^*_0}(x)=\frac{f_{K^*_0}}{2\sqrt{6}},\,\,\,\,\,\,
\int^1_0dx\phi^S_{K^*_0}(x)=\int^1_0dx\phi^{\sigma}_{K^*_0}(x)=\frac{\bar{f}_{K^*_0}}{2\sqrt{6}},
\end{eqnarray}
and
$\phi^T_{K^*_0}(x)=\frac{1}{6}\frac{d}{dx}\phi^{\sigma}_{K^*_0}(x)$.
The twist-2 LCDA can be expanded in the Gegenbauer polynomials:
\begin{eqnarray}
\phi_S(x,\mu)=\frac{1}{\sqrt{2N_c}}\bar
f_S(\mu)6x(1-x)\sum_{m=1}^\infty B_m(\mu)C^{3/2}_m(2x-1).
\end{eqnarray}
The decay constants and the Gegenbauer moments for twist-2 wave
function in two different scenarios  have been studied explicitly in
Ref.~\cite{Cheng:2005ye,Cheng:2005nb} using the QCD sum rule
approach. As for the explicit form of the Gegenbauer moments for the
twist-3 wave functions, there exist few drawbacks in the theoretical
calculation~\cite{Lu:2006fr}, thus we choice the asymptotic form for
simplicity:
\begin{eqnarray}
\phi^s_S= \frac{1}{\sqrt
{2N_c}}\bar f_f,\,\,\,\,\,\,\,\,\,\,\,\,\,\,\,\,\,\,
\phi_S^T=\frac{1}{\sqrt {2N_c}}\bar f_S(1-2x).
\end{eqnarray}
For the pion meson, the asymptotic forms for twist-2 and twist-3
distribution  amplitudes are also adopted:
\begin{eqnarray}
 \phi_P(x)=f_P6x(1-x),\,\,\,\,\,\,\,\,\,
 \phi_P^p(x)=f_P, \,\,\,\,\,\,\,\,\,
 \phi_P^\sigma(x)=f_P6x(1-x).
\end{eqnarray}
The form factors of $B\to P,S$ transitions are defined by \cite{Wirbel:1985ji}:
\begin{eqnarray}
\langle P(p')|V_\mu|B(p)\rangle= \left(P_\mu-{m_B^2-m_P^2\over
q^2}\,q_ \mu\right)
F_1^{BP}(q^2)+{m_B^2-m_P^2\over q^2}q_\mu\,F_0^{BP}(q^2), \non \\
\langle S(p')|A_\mu|B(p)\rangle =
-i\Bigg[\left(P_\mu-{m_B^2-m_S^2\over q^2}\,q_ \mu\right)
F_1^{BS}(q^2) +{m_B^2-m_S^2\over q^2}q_\mu\,F_0^{BS}(q^2)\Bigg],
\end{eqnarray}
where $P_\mu=(p+p')_\mu$, $q_\mu=(p-p')_\mu$. Various form factors
have been evaluated by utilizing the relativistic covariant
light-front quark model~\cite{CCH}. And the momentum dependence is
fitted to a 3-parameter form
\begin{eqnarray}
 F(q^2)=\,{F(0)\over 1-a(q^2/m_{B}^2)+b(q^2/m_{B}^2)^2}\,.
\end{eqnarray}
The parameters $a$ and $b$ relevant for our purposes are refereed to Ref.~\cite{CCH}.

Although we concentrate on the study of new physics, the used
notation for new interacting operators will be similar to those
presented in the SM. Therefore, it is useful to introduce the
effective operators of the SM. Thus, we describe the effective
Hamiltonian for $b\to s q\bar{q}$ decays as
\begin{equation}
H_{{\rm eff}}={\frac{G_{F}}{\sqrt{2}}}\sum_{p=u,c}\lambda_{p}\left[
C_{1}(\mu) O_{1}^{(q)}(\mu )+C_{2}(\mu )O_{2}^{(q)}(\mu
)+\sum_{i=3}^{10}C_{i}(\mu) O_{i}(\mu )\right] \;,
\label{HeffSM}
\end{equation}
where $\lambda_{q}=V_{qb}V_{qs}^{*}$ are the
Cabibbo-Kobayashi-Maskawa (CKM) matrix elements and the operators
$O_{1}$-$O_{10}$ are defined as \cite{Buchalla}
\begin{eqnarray}
&&O_{1}^{(q)}=(\bar{s}_{\alpha}q_{\beta})_{V-A}(\bar{q}_{\beta}b_{\alpha})_{V-A}\;,\;\;\;\;\;
\;\;\;O_{2}^{(q)}=(\bar{s}_{\alpha}q_{\alpha})_{V-A}(\bar{q}_{\beta}b_{\beta})_{V-A}\;,
\nonumber \\
&&O_{3}=(\bar{s}_{\alpha}b_{\alpha})_{V-A}\sum_{q}(\bar{q}_{\beta}q_{\beta})_{V-A}\;,\;\;\;
\;O_{4}=(\bar{s}_{\alpha}b_{\beta})_{V-A}\sum_{q}(\bar{q}_{\beta}q_{\alpha})_{V-A}\;,
\nonumber \\
&&O_{5}=(\bar{s}_{\alpha}b_{\alpha})_{V-A}\sum_{q}(\bar{q}_{\beta}q_{\beta})_{V+A}\;,\;\;\;
\;O_{6}=(\bar{s}_{\alpha}b_{\beta})_{V-A}\sum_{q}(\bar{q}_{\beta}q_{\alpha})_{V+A}\;,
\nonumber \\
&&O_{7}=\frac{3}{2}(\bar{s}_{\alpha}b_{\alpha})_{V-A}\sum_{q}e_{q} (\bar{q}%
_{\beta}q_{\beta})_{V+A}\;,\;\;O_{8}=\frac{3}{2}(\bar{s}_{\alpha}b_{\beta})_{V-A}
\sum_{q}e_{q}(\bar{q}_{\beta}q_{\alpha})_{V+A}\;,  \nonumber \\
&&O_{9}=\frac{3}{2}(\bar{s}_{\alpha}b_{\alpha})_{V-A}\sum_{q}e_{q} (\bar{q}%
_{\beta}q_{\beta})_{V-A}\;,\;\;O_{10}=\frac{3}{2}(\bar{s}_{\alpha}b_{\beta})_{V-A}
\sum_{q}e_{q}(\bar{q}_{\beta}q_{\alpha})_{V-A}\;, \label{eq:ops}
\end{eqnarray}
with $\alpha$ and $\beta$ being the color indices. In
Eq.(\ref{HeffSM}), $O_{1}$-$O_{2}$ are  from the tree level of weak
interactions, $O_{3}$-$O_{6}$ are the so-called QCD penguin
operators and $O_{7}$-$O_{10}$ are the electroweak penguin
operators, while $C_{1}$-$C_{10}$ are the corresponding Wilson
coefficients.

In the QCDF approach, the contribution of the non-perturbative
sector is   dominated by the form factors and the non-factorizable
impact in the hadronic matrix elements is controlled by hard gluon
exchange. The hadronic matrix elements of the decay can be written
as
\begin{eqnarray}
 \langle M_1 M_2|O_i|B\rangle = \sum_{j}F_{j}^{B\rightarrow
M_1 }\int_{0}^{1}dx T_{ij}^{I}(x)\Phi_{M_1}(x)
+\int_{0}^{1}d\xi\int_{0}^{1}dx\int_{0}^{1}dy
T_{i}^{II}(\xi,x,y)\Phi_{B}(\xi)\Phi_{M_1}(x)\Phi_{M_2}(y).
\end{eqnarray}
Here $T_{ij}^{I}$ and $T_{i}^{II}$ denote the perturbative
short-distance  interactions and can be calculated perturbatively.
$\Phi_{X}(x)$ are non-perturbative light-cone distribution
amplitudes, which  should be universal. Using the weak effective
Hamiltonian given by Eq.(\ref{HeffSM}) and the definitions of $a_i$
and $b_i$ in Ref.\cite{Beneke:1999br, Cheng:2005nb}, we can now
write the decay amplitudes of $B\to K_0^*\pi$ as:
\begin{eqnarray}\label{eq:SDAmp}
A(B^- \to \ov K^{*0}_0\pi^- ) &=&
\frac{G_F}{\sqrt{2}}\sum_{p=u,c}\lambda_p
 \Bigg\{ \left( a_4^p-r_\chi^{K^*_0}a_6^p
 -{1\over 2}(a_{10}^p-r_\chi^{K^*_0}a_8^p)\right)_{\pi K^*_0} f_{K_0^*}F_0^{B\pi}(m_{K_0^*}^2)(m_B^2-m_\pi^2)\non \\
 &&
 + f_B\big(b_2\delta_u^p+b_3
 +b_{\rm 3,EW}\big)_{\pi K^*_0} \Bigg\},
\end{eqnarray}
\begin{eqnarray}
A(B^- \to K^{*-}_0\pi^0 ) &=& \frac{G_F}{2}\sum_{p=u,c}\lambda_p
 \Bigg\{ \left( a_1\delta_u^p+a_4^p-r_\chi^{K^*_0}a_6^p
 +a_{10}^p-r_\chi^{K^*_0}a_8^p \right)_{\pi K^*_0}  f_{K_0^*}F_0^{B\pi}(m_{K_0^*}^2)(m_B^2-m_\pi^2)\non \\
 &-&\left[a_2\delta_u^p+{3\over
 2}(a_9-a_7)\right]_{K^*_0\pi}f_\pi F_0^{BK^*_0}(m_\pi^2)(m_B^2-m_{K_0^*}^2)
 +f_B\big(b_2\delta_u^p+b_3
 +b_{\rm 3,EW}\big)_{\pi K^*_0} \Bigg\},\non \\
\end{eqnarray}
\begin{eqnarray}
 A(\ov B^0 \to K^{*-}_0\pi^+ ) &=&
\frac{G_F}{\sqrt{2}}\sum_{p=u,c}\lambda_p
 \Bigg\{ \left( a_1\delta_u^p+ a_4^p-r_\chi^{K^*_0}a_6^p
 +a_{10}^p-r_\chi^{K^*_0}a_8^p \right)_{\pi K^*_0} f_{K_0^*}F_0^{B\pi}(m_{K_0^*}^2)(m_B^2-m_\pi^2) \non \\
 & + &
f_B\big(b_3
 -{1\over 2}b_{\rm 3,EW}\big)_{\pi K^*_0} \Bigg\},
\end{eqnarray}
\begin{eqnarray}
 A(\ov B^0 \to \ov K^{*0}_0\pi^0 ) &=&
\frac{G_F}{2}\sum_{p=u,c}\lambda_p
 \Bigg\{ \left( -a_4^p+r_\chi^{K^*_0}a_6^p
 +{1\over 2}(a_{10}^p-r_\chi^{K^*_0}a_8^p) \right)_{\pi K^*_0}f_{K_0^*}F_0^{B\pi}(m_{K_0^*}^2)(m_B^2-m_\pi^2) \non \\
 &-&
 \left[a_2\delta_u^p+{3\over
 2}(a_9-a_7)\right]_{K^*_0\pi}f_\pi F_0^{BK^*_0}(m_\pi^2)(m_B^2-m_{K_0^*}^2)
 +f_B\big(-b_3
 +{1\over 2}b_{\rm 3,EW}\big)_{\pi K^*_0} \Bigg\}, \non \\
\end{eqnarray}
where $\lambda_p\equiv V_{pb}V_{ps}^*$ and
\begin{eqnarray}\label{eq:r}
  r^{K^*_0}_\chi(\mu)={2m_{K_0^*}^2\over
 m_b(\mu)(m_s(\mu)-m_q(\mu))}.
\end{eqnarray}
In the above formulaes, the order of the arguments of the
$a_i^p(M_1M_2)$ and  $b_i(M_1M_2)$  coefficients is dictated by the
subscript $M_1M_2$, where $M_2$ is the emitted meson and $M_1$
shares the same spectator quark with the $B$ meson. For the
annihilation diagram, $M_1$ is referred to the one containing an
anti-quark from the weak vertex, while $M_2$ contains a quark from
the weak vertex. Note that the coefficients $a_i$ come from vertex
corrections and hard  spectator corrections, and $b_i$ represent of
contribution of annihilation diagrams. Both $a_i$ and $b_i$ can be found in
Ref.\cite{Cheng:2005nb}. It must be emphasized that we shall
evaluate the vertex corrections to the decay amplitudes at the scale
$\mu=m_b/2$. In contrast, the hard spectator and annihilation
contributions should be evaluated at the hard-collinear scale
$\mu_h=\sqrt{\mu\Lambda_h}$ with $\Lambda_h\approx 500 $ MeV.

In QCDF approach, the annihilation amplitude has endpoint
divergences even at  twist-2 level and the hard spectator scattering
diagram at twist-3 order is power suppressed and posses soft and
collinear divergences arising from the soft spectator quark. Since
the treatment of endpoint divergences is model dependent,
subleading power corrections generally can be studied only in a
phenomenological way. We shall follow
\cite{Beneke:1999br,Cheng:2005nb} to parameterize the endpoint
divergence $X_A\equiv\int^1_0 dx/\bar x$ in the annihilation diagram
as
\begin{eqnarray} \label{eq:XA}
 X_A=\ln\left({m_B\over \Lambda_h}\right)(1+\rho_A e^{i\phi_A}),
\end{eqnarray}
with the unknown real parameters $\rho_A$ and $\phi_A$. Likewise,
the endpoint divergence $X_H$ in the hard spectator contributions
can be parameterized in a similar manner. In the Sec.\ref{sec:4}, we
will see that such divergence is  the main source of the uncertainty
for the concerned decay modes.

\section{The Family Non-universal $Z^\prime$ Model}\label{sec:3}
As mentioned before, a family non-universal $Z^{\prime}$ model leads
to FCNC at the tree level due to the non-diagonal chiral coupling
matrix, which makes itself become interesting in some penguin
dominate processes. The basic formalism of flavor changing effects
in the $Z^\prime$ model with family  nonuniversal and/or nondiagonal
couplings has been laid out in
Refs.~\cite{Langacker:2000ju,Langacker:2008yv}, to which we refer
readers for  detail. The detailed phenomenological analysis for
various low energy physics, especially for $B$ meson decays, could
be found in Refs.~\cite{Barger:2009hn,Cheung:2006tm,Chang:2009wt}.
Here we just briefly review the ingredients needed in this paper.

In practice, neglecting the renormalization group (RG) running
between $m_W$ and $m_{Z^\prime}$ and mixing between $Z^\prime$  and
$Z$ boson of the SM, we write the $Z^\prime$ term of the neutral-current Lagrangian in the gauge basis as
\begin{eqnarray}
 {\cal L}=-g^\prime J^\prime_\mu Z^{\prime \mu},
\end{eqnarray}
where $g^\prime$ is the gauge coupling constant of extra
$U(1)^\prime$ group at  the electro-weak $m_W$ scale. The chiral
current $J^{\prime}_\mu$ is expressed as:
\begin{eqnarray}
J^\prime_\mu=\bar
\psi_i\gamma_\mu\bigg[(B_{ij}^L)P_L+(B_{ij}^R)P_R\bigg]~\psi_j,
\end{eqnarray}
where the chirality projection operators are
$P_{L,R}\equiv(1\pm\gamma_5)/2$ and   $B_{ij}^X$ refers to the
effective $Z^{\prime}$ couplings to the quarks $i$ and $j$ at the
electroweak scale. For simplicity, we assume that the right hand
couplings are flavor-diagonal and  neglect $B_{sb}^R$. Compared with
Eq.(\ref{HeffSM}), the effective Hamiltonian for $b\to s\bar{q}q$
transition with $Z^\prime$ boson can be written as
\begin{eqnarray}\label{heffz1}
 {\cal H}_{\rm eff}^{\rm
 Z^{\prime}}=\frac{2G_F}{\sqrt{2}}\big(\frac{g^{\prime}m_Z}
 {g_1m_{Z^{\prime}}}\big)^2
 \,B_{sb}^L(\bar{s}b)_{V-A}\sum_{q}\big[B_{qq}^L (\bar{q}q)_{V-A}
+B_{qq}^R(\bar{q}q)_{V+A}\big]+{\rm h.c.}\,,
\end{eqnarray}
where $m_{Z^{\prime}}$ is the mass of the new gauge boson. In fact,
the forms of four-quark operators in Eq.~(\ref{heffz1}) already
exist in the SM, so we rewrite it  as
\begin{eqnarray}
 {\cal H}_{\rm eff}^{\rm
 Z^{\prime}}=-\frac{G_F}{\sqrt{2}}V_{tb}V_{ts}^{\ast}\sum_{q}
 \big(\Delta C_3 O_3^q +\Delta C_5 O_5^q +\Delta C_7 O_7^q+\Delta C_9
  O_9^q\big)+{\rm h.c.}\,,
\end{eqnarray}
where $O_i^q~(i=3,5,7,9)$ are the effective four-quark operators in
the SM.  $\Delta C_i$ denote the modifications to the corresponding
SM Wilson coefficients, which are expressed as
\begin{eqnarray}
 \Delta C_{3,5}&=&-\frac{2}{3V_{tb}V_{ts}^{\ast}}\,\big(\frac{g^{\prime}m_Z}
 {g_1m_{Z^{\prime}}}\big)^2\,B_{sb}^L\,(B_{uu}^{L,R}+2B_{dd}^{L,R})\,,\nonumber\\
 \Delta C_{9,7}&=&-\frac{4}{3V_{tb}V_{ts}^{\ast}}\,\big(\frac{g^{\prime}m_Z}
 {g_1m_{Z^{\prime}}}\big)^2\,B_{sb}^L\,(B_{uu}^{L,R}-B_{dd}^{L,R})\,,
 \label{NPWilson}
\end{eqnarray}
Generally, the diagonal elements of the effective coupling matrices
$B_{qq}^{L,R}$  are expected to be real as a consequence of the
hermiticity of the effective weak Hamiltonian. However, the
off-diagonal one  $B_{sb}$ perhaps contains a new weak phase
$\phi_s$. We also suppose $B_{qq}^{L} =B_{qq}^{R} =B_{qq}$, so as to
reduce the new parameters. For convenience we can represent $\Delta
C_i$ as
\begin{eqnarray}\label{deltac3597}
 \Delta
 C_{3,5}=2\,\frac{|V_{tb}V_{ts}^{\ast}|}{V_{tb}V_{ts}^{\ast}}\,
 \zeta\,e^{i\phi_s}\,,\quad
 \Delta
 C_{9,7}=4\,\frac{|V_{tb}V_{ts}^{\ast}|}{V_{tb}V_{ts}^{\ast}}\,
 \xi\,e^{i\phi_s}\,,
\end{eqnarray}
where  $\zeta$ and $\xi$ are defined, respectively, as
\begin{eqnarray}\label{zetaandxi}
 \zeta&=&-\frac{1}{3}\,\big(\frac{g^{\prime}M_Z}
 {g_1M_{Z^{\prime}}}\big)^2\,\big|\frac{B_{sb}^L}{V_{tb}V_{ts}^{\ast}}\big|\,
 (B_{uu}+2B_{dd}),\non\\
 \xi&=&-\frac{1}{3}\,\big(\frac{g^{\prime}M_Z}{g_1M_{Z^{\prime}}}\big)^2\,
 \big|\frac{B_{sb}^L}{V_{tb}V_{ts}^{\ast}}\big|\,(B_{uu}-B_{dd}).
\end{eqnarray}
It is stressed that the other SM Wilson coefficients may also
receive contributions  from the $Z^{\prime}$ boson through
renormalization group~(RG) evolution. With our assumption that no
significant RG running effect between $M_Z^{\prime}$ and $M_W$
scales, the RG evolution of the modified Wilson coefficients is
exactly the same as the ones in the SM~\cite{Buchalla}.

In order to show the effects of $Z^\prime$ boson clearly, our
analysis are divided into the two cases with two different
simplifications,
\begin{eqnarray}
\left\{
  \begin{array}{llll}
    B_{uu}=-2B_{dd},&\zeta=0,&\xi=X, & \hbox{Case I;} \\
    B_{uu}= B_{dd},&\zeta=-X,&\xi=0, & \hbox{Case II.}
  \end{array}
\right.
\end{eqnarray}
with
\begin{eqnarray}
X=\big(\frac{g^{\prime}M_Z}{g_1M_{Z^{\prime}}}\big)^2\,
 \left|\frac{B_{sb}^LB_{dd}}{V_{tb}V_{ts}^{\ast}}\right|
 =y\left|\frac{B_{sb}^LB_{dd}}{V_{tb}V_{ts}^{\ast}}\right|\,.
\end{eqnarray}
Thus, there are only two parameters, $X$ and weak phase $\phi_s$
left, in the sequential  numerical calculations and discussions.

\section{Numerical Results and Discussions}\label{sec:4}

To obtain the numerical results, we list the parameters related to the
SM firstly.  As stated in Section.~\ref{sec:1}, because we have not
a clear conclusion whether  $K_0^*(1430)$ belongs to the first
orbitally excited state (S1) or the low lying state (S2), we have to
calculate the processes under both scenarios. So, the decay
constants, Gegenbauer moments, and form factors in different
scenarios are listed as follows \cite{Cheng:2005nb}:
\begin{eqnarray}
 &\mathbf{ S1}: \bar f_{K_0^*}(1.0~~\mathrm{GeV})=-300 ~~\mathrm{MeV};
       \bar f_{K_0^*}(2.1~~\mathrm{GeV})=-370~~\mathrm{MeV};
      B_1(1.0~~\mathrm{GeV})=0.58;\non\\
 &
       B_1(2.1~~\mathrm{GeV})=0.39;     B_3(1.0~~\mathrm{GeV})=-1.20;
      B_3(2.1~~\mathrm{GeV})=-0.70;
      F_0^{BK_0^*}(0)=F_1^{BK_0^*}(0)=0.21;  \\
 &\mathbf{ S2} : \bar f_{K_0^*(1430)}(1.0~~\mathrm{GeV})=445 ~~ \mathrm{MeV};
     \bar f_{K_0^*(1430)}(2.1~~\mathrm{GeV})=550~~\mathrm{MeV};
      B_1(1.0~~\mathrm{GeV})=-0.57;\non\\
&
      B_1(2.1~~\mathrm{GeV})=-0.39;    B_3(1.0~~\mathrm{GeV})=-0.42;
      B_3(2.1~~\mathrm{GeV})=-0.25;
      F_0^{BK_0^*}(0)=F_1^{BK_0^*}(0)=0.26;
\end{eqnarray}
Now that the uncertainties for the above parameters have been
explored explicitly in Ref.\cite{Cheng:2005nb}, and we will not
discuss the errors caused by them in the current work.

In Ref.~\cite{Cheng:2005nb}, the authors concluded that the
theoretical errors are  dominated by the $1/m_b$ power corrections
due to the weak annihilations. Moreover, the weak annihilation
contributions to $B\to SP$ could be much larger than the $B\to PP$
case, because the helicity suppression appeared in the $B\to PP$
case can be alleviated in the scalar production with the
non-vanishing orbital angular momentum in the scalar state. In order
to accommodate the data, one has to take into account the power
corrections due to the $\rho_H$ and $\rho_A$ from the hard spectator
interactions and weak annihilations, respectively. In Ref.
\cite{Cheng:2005nb}, Cheng {\it et.al} found that the predictions
are far away from the experimental data if by setting $\rho_A=0$,
which indicates that $\rho_A$ will be nonzero. Meanwhile, for $B \to
PP,PV$ modes \cite{Beneke:1999br}, the errors due to weak
annihilations are comparable to or much smaller than the center
values, and the fitting results show that $\rho_A=1$ and
$\phi_A=0^\circ$. Hence, in this work, we  adopt $\rho_H=\rho_A=1$,
and set the strong phases $\phi_{A,H}$ in the ranges $[-30^\circ, 30
^\circ]$.

\begin{table*}
\begin{center}
\caption{Branching ratios (in units of $10^{-6}$) of $B\to K_0^*\pi$
in the SM and the non-universal $Z'$ model.}\label{Table:1}
\begin{tabular}{c|ccc|ccc|c }
\hline\hline
 &\multicolumn{3}{|c|}{S1}& \multicolumn{3}{|c|}{S2}& \\
Decay Mode  &SM  & Case I  & Case II & SM  &  Case  I &   Case  II&  Expt \\
\hline $B^-\to \overline K_0^{*0}\pi^-$ &$23.0^{+1.2}_{-5.9}$
&$25.7^{+5.0+2.8}_{-4.7-7.7}$ &$17.2^{+1.3+19.6}_{-5.1-5.0}$
&$74.7^{+1.0}_{-20.6}$ &$93.8^{+1.7+20.6}_{-25.8-50.6}$
&$53.8^{+1.3+16.5}_{-17.5-17.4}$ &$45.1^{+6.3}_{-6.3}$
\\
$B^-\to K_0^{*-}\pi^0$ &$ 9.3^{+1.0}_{-1.9}$
&$17.9^{+0.8+11.4}_{-3.3-17.3}$ &$6.8^{+1.1+8.7}_{-1.6-2.2}$
&$38.9^{+0.4}_{-8.9}$ &$75.3^{+0.0+46.8}_{-14.5-73.1}$
&$28.2^{+0.5+8.4}_{-7.7-8.9}$ &$ $
\\
$\overline B^0\to K_0^{*-}\pi^+$ &$21.3^{+0.7}_{-5.1}$
&$27.2^{+0.2+6.8}_{-6.4-15.6}$ &$15.9^{+0.8+18.4}_{-4.4-4.6}$
&$70.0^{+0.6}_{-17.2}$ &$83.3^{+0.0+13.4}_{-16.2-36.3}$
&$50.2^{+0.9+15.6}_{-14.6-16.4}$ &$33.5^{+3.9}_{-3.8}$
\\
$\overline B^0\to\overline K_0^{*0}\pi^0$ &$12.9^{+0.3}_{-3.6}$
&$9.4^{+0.3+12.0}_{-2.7-2.9}$ &$9.8^{+0.3+10.3}_{-3.1-2.7}$
&$33.6^{+0.4}_{-9.8}$ &$22.0^{+0.5+41.9}_{-7.2-8.5}$
&$23.8^{+0.5+7.7}_{-8.2-8.0}$
&$11.7^{+4.2}_{-3.8}$\\
\hline\hline
\end{tabular}
\end{center}
\end{table*}
\begin{table*}
\begin{center} \caption{$CP$ asymmetry (in $\%$) of $B\to K_0^*\pi$
in the SM and the non-universal $Z'$ model.}\label{Table:2}
\begin{tabular}{c|ccc|ccc|c }
\hline\hline
 &\multicolumn{3}{|c|}{S1}& \multicolumn{3}{|c|}{S2} \\
Decay Mode  &SM  & Case I  & Case II & SM  &  Case  I &   Case  II & Expt\\
\hline $B^-\to \overline K_0^{*0}\pi^-$ &$1.0^{+1.9}_{-1.9}$
&$1.0^{+1.4+0.2}_{-1.6-0.1}$ &$1.2^{+2.4+0.2}_{-2.1-0.4}$
&$0.06^{+0.6}_{-0.7}$ &$0.06^{+0.4+0.3}_{-0.5-0.3}$
&$0.03^{+0.66+0.03}_{-0.82-0.07}$ &$-5^{+5}_{-8}$
\\
$B^-\to K_0^{*-}\pi^0$ &$-0.5^{+3.8}_{-2.6}$
&$-0.4^{+2.7+0.6}_{-2.0-2.8}$ &$-0.6^{+4.6+0.4}_{-2.8-0.2}$
&$1.0^{+2.4}_{-2.9}$ &$0.7^{+1.8+5.4}_{-2.0-1.2}$
&$1.2^{+2.7+0.3}_{-3.5-0.2}$ &$ $
\\
$\overline B^0\to K_0^{*-}\pi^+$ &$2.0^{+2.7}_{-3.9}$
&$1.7^{+2.3+1.5}_{-2.6-0.7}$ &$2.4^{+3.1+1.0}_{-4.6-1.4}$
&$-0.8^{+2.9}_{-2.7}$ &$-0.7^{+2.1+0.6}_{-2.2-0.9}$
&$-1.1^{+3.6+0.2}_{-3.3-0.5}$ &$-7^{+14}_{-14}$
\\
$\overline B^0\to\overline K_0^{*0}\pi^0$ &$3.1^{+2.6}_{-4.0}$
&$3.7^{+3.0+1.3}_{-4.7-2.0}$ &$3.7^{+2.9+1.0}_{-4.5-1.8}$
&$-1.9^{+4.0}_{-3.5}$ &$-2.5^{+5.3+3.3}_{-4.4-2.2}$
&$-2.5^{+4.9+0.5}_{-4.1-1.0}$ &$-34^{+19}_{-19}$
\\
\hline\hline
\end{tabular}
\end{center}
\end{table*}

With above parameters, we  present our predictions of the SM in
Table.\ref{Table:1} under two different scenarios. For the center
values, we also assign $\phi_A=\phi_H=0$. In order to obtain the
errors, we scan randomly the points in the ranges $\phi_A
\in[-30^\circ,30^\circ]$ and $\phi_H \in[-30^\circ, 30^\circ]$. So,
the only theoretical errors of the SM results are due to the strong
phases $\phi_A$ and $\phi_H$. Because we fully consider the weak
annihilations, our results are much larger than those in
Ref.\cite{Cheng:2005nb}, especially for the center values. Compared
with the data, the theoretical results in this work are still much
smaller (larger) than the data under two scenarios, except for mode
$\overline B^0\to\overline K_0^{*0}\pi^0$. If one want to fit the
data absolutely, $\rho_A\approx1.3$ for S1 and $\rho_A\approx0.7$
for S2 are required, respectively, which are a bit larger/smaller by
$30\%$ than the fitted results from $B \to PP, PV$. Compared with
predictions of Ref.~\cite{shen} obtained in the pQCD approach based
on $k_T$ factorization, our results are a bit larger than theirs in
S2, but agree with their results in S1 with large uncertainties.

We next turn to the implications of the non-universal $Z^\prime$
model for the $B \to K_0^*\pi$ decays. Let us firstly consider the
range of $X$, which is the most important parameter in this model.
Generally, we always expect $g^\prime/g_1 \sim 1 $, if both the
$U(1)$ gauge groups have the same origin from some grand unified
theories.  $M_Z/M_{Z^\prime} \sim 0.1$ for TeV scale neutral
$Z^\prime$ boson is also expected so as to the $Z^\prime$ could be
detected in the running Large Hadron Collider (LHC), which results
in $y \sim 10^{-2}$. In the first paper of Ref. \cite{Barger:2009hn}
assuming a small mixing between $Z-Z^\prime$ bosons, the value of
$y$ is taken as $y \sim 10^{-3}$. In order to explain the mass
difference of $B_s -\bar B_s $ mixing, $|B_{sb}^L|\sim |V_{tb}
V_{ts}^*| $ is required. Similarly, the CP asymmetries in $B \to
\phi K, \pi K $ can be resolved if $|B_{sb}^L B_{ss}^{L,R}| \sim
|V_{tb} V_{ts}^*|$, which indicates $|B_{qq}^L|\sim 1 $. Above
issues have been discussed widely in Ref. \cite{Cheung:2006tm}.
Summing up above analysis, we thereby assume that $X\in(10^{-3},
10^{-2})$. For weak phase $\phi_s$, though many attempts have been
done to constrain it \cite{Chang:2009wt}, we here left it as a free
parameter.

The calculated results for branching ratios with two different cases
in the family non-universal $Z^\prime$ model are also exhibited in
Table.\ref{Table:1}, and  for the center values we use $X=0.005$ and
$\phi_s=0^\circ$. To obtain the second errors, we also scan randomly
the points in the ranges $X \in[0.001,0.01]$ and $\phi_s
\in[-180^\circ,180^\circ]$, while the first errors come from the
weak annihilations. The table shows to us that the two cases of
$Z^\prime$ models can change the branching ratios remarkably in the
two different scenarios. It is clear that the $Z^\prime$ will
enhance the branching ratios in Case-I, while in Case-II the
branching ratios are decreased. The reason is that the variation
tendencies of Wilson coefficients are different in the two different
cases, which could be seen in Eq.(25) and (27) easily.

For S1, the branching ratios of the first three decay modes cannot
agree with data unless the upper limits in Case-II of the $Z^\prime$
model are taken. Unfortunately, with the upper limit values, the
branching ratio of $\overline B^0\to\overline K_0^{*0}\pi^0$ is much
larger than the experimental data. For S2, the branching ratios with
a  $Z^\prime$ boson can accommodate experimental data well in two
cases with large uncertainties. If we care about the center values
very much, it seems that results of Case-II are preferable. If
further theories and/or experiments  can confirm the existence of
$Z^\prime$, one could correspondingly cross-check the couplings and
the mass of it with all above results in turn.

In the experimental side, another important observable in $B$
physics is $CP$ asymmetry, in particular of the direct $CP$
asymmetry. In Table.\ref{Table:2}, we list the direct $CP$
asymmetries of concerned modes in different scenarios and different
cases of the $Z^\prime$ model. Generally, the strong phases
calculable in the QCD factorization are so small that the $CP$
asymmetries are at most a few percent, as shown in the table. In S1
we note that the center values have different signs with the
experimental data. Adding the $Z^\prime$ contribution, although the
large uncertainties perhaps alleviate the disparity of $B^-\to
\overline K_0^{*0}\pi^-$, but the large asymmetries in the
$\overline B^0\to K_0^{*-}\pi^+$ and $\overline B^0\to
K_0^{*0}\pi^0$ cannot be explained yet. In S2, for $\overline B^0\to
K_0^{*-}\pi^+$ and $\overline B^0\to K_0^{*0}\pi^0$, the signs of
center values are same as those of data. Furthermore, the $CP$
asymmetries of $B^-\to \overline K_0^{*0}\pi^-$ in the SM and
$Z^\prime$ model are almost null, which are close to the upper
limits of experiment. Considering the large uncertainties, the
results of S2 in both SM and $Z^\prime$ models can accommodate the
data, except for the unexpectedly large asymmetry of $\overline
B^0\to K_0^{*0}\pi^0$, which should be measured critically in
future. However, as pointed out in Ref. \cite{CCS}, final state
interaction may have important effects on the decay rates and their
direct $CP$ violations, especially for the latter. However, this is
beyond the scope of the present work.

\begin{figure*}[t]
  \centerline{\psfig{figure=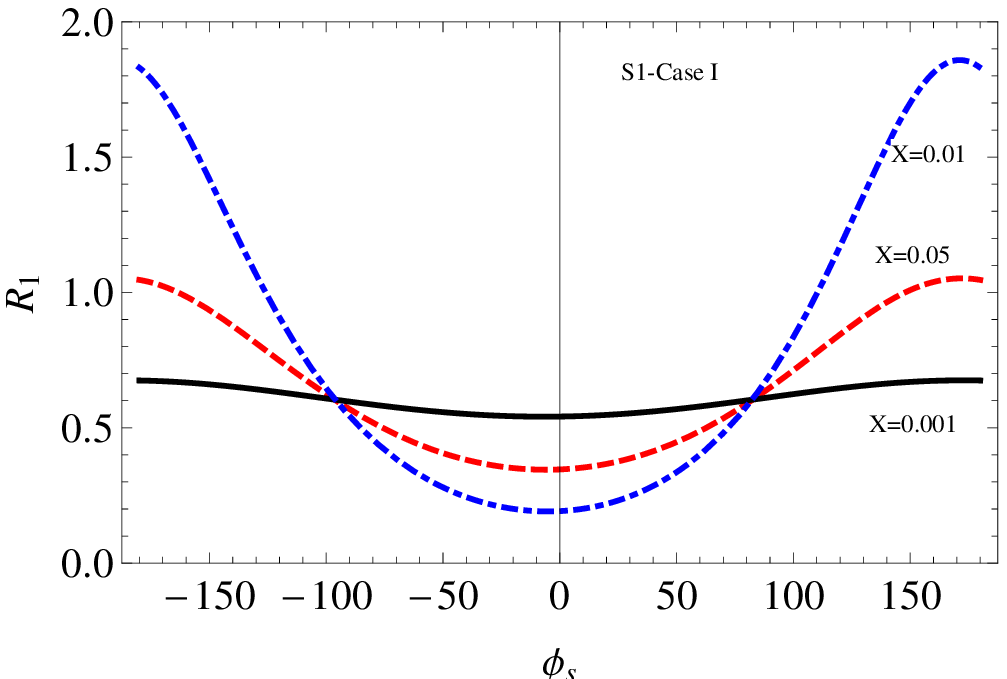,width=7cm}\,\,\,\,\,\,\,\,\,\,\,\,\,\,\,\,\,\,
  \psfig{figure=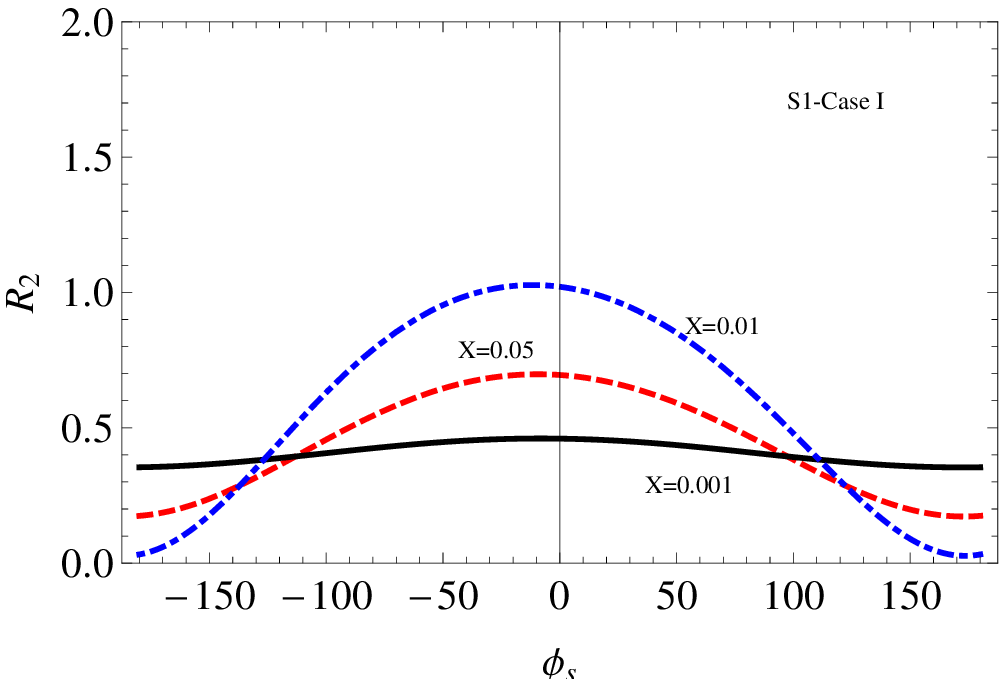,width=7cm}}
  \vspace{1.0cm}
 \centerline{\psfig{figure=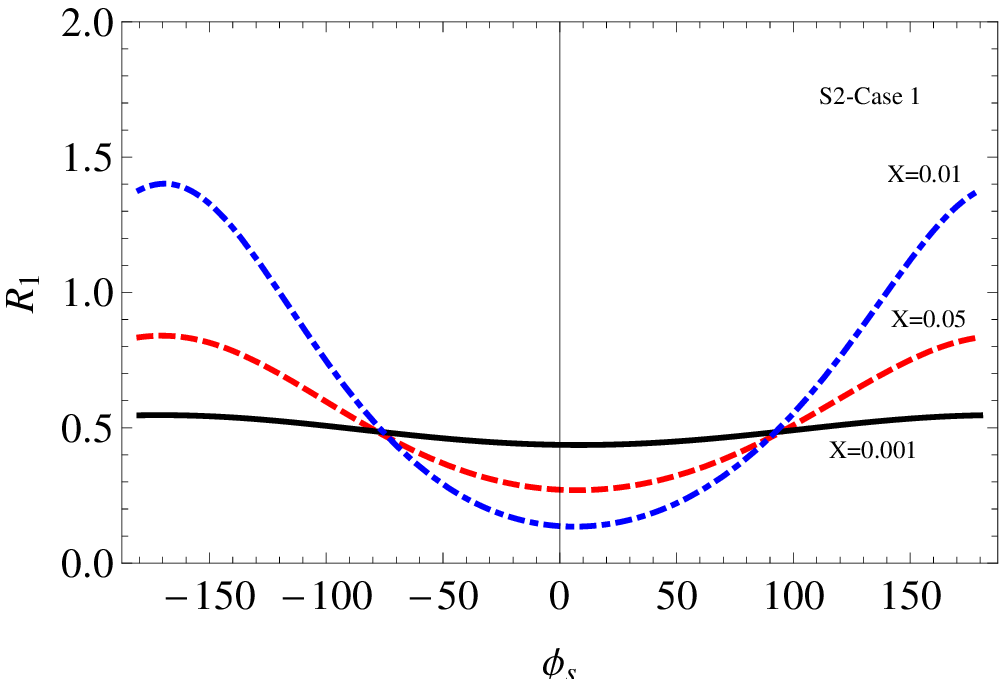,width=7cm}\,\,\,\,\,\,\,\,\,\,\,\,\,\,\,\,\,\,
  \psfig{figure=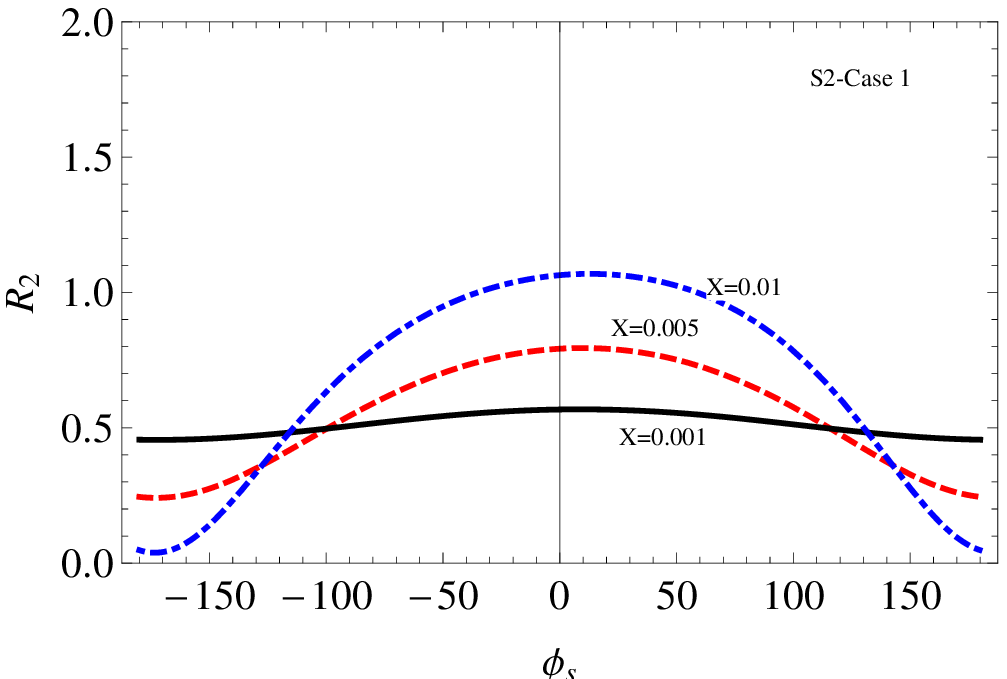,width=7cm}}
    \caption{$R_1$ and $R_2$ as functions of weak phase $\phi_s$ with different $X$ in different scenarios and cases.}
     \label{fig:1}
\end{figure*}

Let us now analyze the impact of $Z^\prime$ on the isospin symmetry
breaking. To explore the deviation from the isospin limit, it is
convenient to define the following three parameters:
\begin{eqnarray} \label{eq:ratios}
 R_1&\equiv& {\B(\ov B^0\to \ov K_0^{*0}(1430)\pi^0)\over \B(\ov B^0\to
 K_0^{*-}(1430)\pi^+)} ,  \\
 R_2&\equiv& {\B(B^-\to K_0^{*-}(1430)\pi^0)\over \B(B^-\to \ov
 K_0^{*0}(1430)\pi^-)},   \\
 R_3&\equiv&\frac{\tau(B^0)}{\tau(B^-)} \,\,{\B(B^-\to \ov K_0^{*0}(1430)\pi^-)\over \B(\ov B^0\to
 K_0^{*-}(1430)\pi^+)}.
\end{eqnarray}
Because they are the ratios of the branching fractions, they should
be less sensitive to the non-perturbative inputs  than other
observables discussed before, therefore it is more persuasive to
test them in both theoretical and experimental sides. In the isospin
limits, i.e., ignoring the electroweak penguins, $R_1$, $R_2$ and
$R_3$  are equal to 0.5, 0.5, and 1.0, respectively. So, the
deviations reflect the magnitudes of the electroweak penguins
directly. The results of SM and the non-universal $Z^\prime$ model
are listed in Table.\ref{Table:3}. In the SM, it appears that the
deviations from the isospin limit are not large in both scenarios,
which shows that the QCD penguins are dominant. For Case-I of the
$Z^\prime$ model, the new physics just revise the Wilson
coefficients of electroweak penguin operators, which could break the
isospin symmetry. So, the ratios will be changed remarkably in both
scenarios, as shown in the table. In Figure.\ref{fig:1}, we also
present the variations $R_{1,2}$ as functions of the new weak phase
$\phi$ with different $X=0.001, 0.005, 0.01$ in S1 (up panels) and
S2 (down panels), so as to show the effect of two parameters $X$ and
$\phi$. From the figures, we see that the $R_{1,2}$ change
remarkably when $X=0.01$ and $0.005$. As $X=0.001$, $R_{1,2}$ almost
have same values as predictions of the SM. For Case-II, the
$Z^\prime$ boson changes the Wilson coefficients of QCD penguins, so
the isospin symmetries are almost unchanged, as shown in
Table.\ref{Table:3}. To sum up, the measurements of the $R_i$ will
help us determine whether QCD or electroweak interactions will be
changed and then test the corresponding new physics models.

\begin{table*}
\begin{center}
 \caption{Ratios of the branching fractions in the SM and the non-universal $Z'$ model.}\label{Table:3}
\begin{tabular}{c|ccc|ccc|c }
\hline\hline
 &\multicolumn{3}{|c|}{S1}& \multicolumn{3}{|c|}{S2} \\
$~~~~R_i~~~~$  &SM  & Case I  & Case II & SM  &  Case  I &   Case  II&   Expt. \\
\hline $R_1$ &$0.60^{+0.00}_{-0.04}$
&$0.35^{+0.03+1.50}_{-0.02-0.15}$ &$0.62^{+0.00+0.02}_{-0.05-0.03}$
&$0.48^{+0.01}_{-0.03}$ &$0.26^{+0.02+1.09}_{-0.04-0.12}$
&$0.48^{+0.01+0.00}_{-0.03-0.00}$ &$0.35^{+0.18}_{-0.15}$
\\
$R_2$ &$0.40^{+0.04}_{-0.00}$ &$0.70^{+0.04+0.33}_{-0.09-0.66}$
&$0.39^{+0.05+0.02}_{-0.00-0.01}$ &$0.52^{+0.03}_{-0.01}$
&$0.80^{+0.09+0.26}_{-0.05-0.75}$ &$0.52^{+0.04+0.01}_{-0.01-0.00}$
&
\\
$R_3$ &$1.00^{+0.03}_{-0.02}$ &$0.87^{+0.28+0.59}_{-0.05-0.10}$
&$1.00^{+0.04+0.01}_{-0.02-0.01}$ &$0.99^{+0.02}_{-0.03}$
&$1.05^{+0.08+0.27}_{-0.10-0.21}$ &$0.99^{+0.02+0.01}_{-0.04-0.00}$
&$1.25^{+0.36}_{-0.29}$
\\
\hline \hline
\end{tabular}
\end{center}
\end{table*}

Finally, we will go back to the discussion of two scenarios. As
aforementioned, $K_0^*(1430)$ is regarded as two-quark state in both
S1 and S2, but the only controversy is whether it belongs to ground
state or the first excited state. Through calculation and comparison
above, we favor the second scenario, which means that $K_0^*(1430)$
is the lowest lying $\bar qq$ state. Namely, the scalar mesons lower
than $1 \mathbf{GeV}$ are four-quark states. This conclusion is also
consistent with those of Refs.\cite{Cheng:2005nb, shen,
Chen:2002si}.

\section{Summary}\label{sec:5}
Based on the  QCD factorization approach, we have investigated in
this work $B \to K_0^*\pi$ decays  in the SM and a family
non-universal $Z^\prime$ model. Because the inner structure of
$K_0^*(1430)$ is not clear enough, we calculated the branching
ratios under two different scenarios (S1 and S2).  After
calculation, we found that the branching ratios are sensitive to the
weak annihilations. In the SM, with $\rho_A=1$ and
$\phi_A\in[-30^\circ, 30^\circ]$, the branching ratios of S1 (S2)
are smaller (larger) than the experimental data. Considering the
$Z^\prime$ boson in two different cases, for S1, the branching
ratios are still far away from experiment. For S2, the branching
ratios become smaller and can accommodate the data  in Case-II; in
Case-I, the results can also explain the data but with large
uncertainties. Furthermore, the other interesting observables, such
as $CP$ asymmetries and isospin asymmetries, are also calculated.
Compared with data, we favor that $K_0^*(1430)$ is the lowest lying
$\bar qq$ state. Moreover, if there exists a $Z^\prime$ boson,
Case-II is preferable. All above results will be tested in the B
factories, LHC-b and the forthcoming super-B factory.

\section*{Acknowledgement}
The work of Y.Li was supported by the National Science Foundation
(Nos.10805037 and 11175151) and the Natural Science Foundation of
Shandong Province (ZR2010AM036). Y.Li also thanks Hai-Yang Cheng and 
Kwei-Chou Yang for useful comments.


\end{document}